\documentclass[twocolumn,aps,superscriptaddress,prb,longbibliography,preprintnumbers]{revtex4-1}

\setcitestyle{numbers,square}
\usepackage{amsmath,amssymb}
\usepackage{tabularx}
\usepackage{graphicx}
\usepackage{bmpsize}
\usepackage{bm}
\usepackage{color}
\usepackage{amssymb}
\usepackage[version=3]{mhchem}
\usepackage{newtxmath}
\usepackage{physics}
\usepackage{hyperref}
\usepackage{mathtools}
\usepackage{xcolor}
\usepackage{ulem}
\usepackage{booktabs}  
\usepackage{makecell}
\newcolumntype{Y}{>{\centering\arraybackslash}X}
\DeclareMathAlphabet{\mathpzc}{OT1}{pzc}{m}{it}

\def\diag{\mathrm {diag}}

\begin{document}
\preprint{OU-HET-1280}
\title{Time-reversal invariant vortex in topological superconductors and gravitational $\mathbb{Z}_2$ topology}
\author{Kazuki Yamamoto}
\author{Naoto Kan}
\author{Hidenori Fukaya}
\affiliation{Department of Physics, The University of Osaka, Toyonaka, Osaka 560-0043, Japan}


\begin{abstract}
We study 
a time-reversal invariant vortex, namely a spin vortex, in helical superconductors
by focusing on its emergent gravitational structure.
The topology of the time-reversal invariant vortex is classified by a $\mathbb{Z}_2$ invariant:
a zero-energy Majorana Kramers pair appears at the vortex core when the winding number is odd, while no such zero modes exist when it is even.
We provide a formal mapping to the theory of gravity
to describe this $\mathbb{Z}_2$ topological structure.
Identifying a superconducting order parameter as a vielbein in the theory of gravity, we explicitly convert the Bogoliubov-de-Genne (BdG) Hamiltonian into the Dirac Hamiltonian coupled to a nontrivial gravitational field.
Then we find that a gravitational curvature is induced at the vortex core, with its total flux quantized in integer multiples of $\pi$, reflecting the $\mathbb{Z}_2$ topology.
Although the curvature vanishes everywhere except at the vortex core, the energy spectrum remains sensitive to the total curvature flux, owing to the gravitational Aharonov-Bohm effect.
We further demonstrate that our gravitational framework can be applied to the topological phase transition driven by the vortex-linking precess in three-dimensional helical superconductors such as the He-B phase.
\end{abstract}
\maketitle

\section{Introduction}
\begin{figure}[b]
    \centering
    \includegraphics{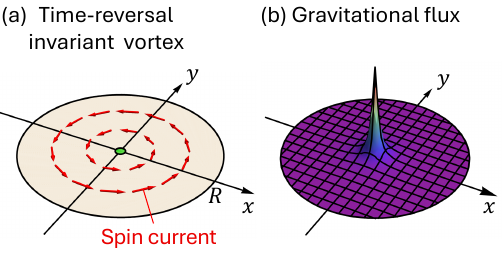}
    \caption{
    Schematic illustration of (a) a time-reversal invariant vortex in a two-dimensional helical superconductor and (b) the corresponding induced gravitational flux. 
    }
    \label{figure:schematic}
\end{figure}
Time-reversal invariant topological superconductors belong to a class of superconducting materials characterized by a topological full pairing gap in the bulk and symmetry-protected helical edge-localized states at the boundaries~\cite{ReadGreen2000,Qi2009,Sato2017}.
These systems preserve time-reversal symmetry and are topologically distinct from conventional superconductors.

A hallmark feature of such phases is possible emergence of the zero-energy Majorana Kramers pair~\cite{Zhang_PRL2013,Wolms_PRB2016,Li_PRL2016,Hsu_PRL2018,Kobayashi_PRL2019,Yamazaki_JPSJ2020,Yamazaki_PRB2021,Yamazaki_JPSJ2021,Schrade_PRL2022} at the edge.
Being a Kramers pair, they are topologically protected by time-reversal symmetry.
Majorana fermions are exotic particles characterized by their invariance under particle-hole transformation, meaning that they are their own antiparticles.
Although Majorana fermions were originally predicted in particle physics, their existence has not yet been confirmed experimentally. 

Previous studies on the possible zero-energy Majorana Kramers pair have mainly focused on edge states, localized at surfaces of dimension ($d-1$), where $d$ is the spatial dimension of the bulk system. 
An interesting question is whether or not 
the zero-energy Majorana Kramers pair
can appear in the lower-dimensional defects in the superconductors.
One often considers a U(1) vortex in type II superconductors, that is a ($d-2$)-dimensional defect.
However, these conventional U(1) vortices carry magnetic flux localized at their core, which inherently breaks the time-reversal symmetry, which precludes the presence of the zero-energy Majorana Kramers pair at the vortex core.

To overcome this limitation, we instead consider a time-reversal invariant vortex~\cite{Qi2009}. 
Unlike conventional vortices that wind the U(1) phase of the order parameter, the time-reversal invariant vortex involves a winding of the spin degrees of freedom (i.e. SO(2)),
leading to the circulation of the spin supercurrent around the vortex.
It enables the realization of the zero-energy Majorana Kramers pair without breaking time-reversal symmetry,
because, in the latter case, the order parameter remains  real.

While conventional U(1) vortices in type-II superconductors can be spatially regulated through an external magnetic field, the time-reversal invarinat vortex (i.e. pure spin vortex) lack such a controllable external field, rendering their experimental investigation challenging.
(Note that,
in Ref.~\cite{Volovik2009}, they proposed a realistic setup to observe a spin vortex coexisting with a conventional mass vortex in the He-B phase. However, this is not the case we consider, since our study focuses on the time-reversal invariant vortex, namely a pure spin vortex without an accompanying mass vortex.)

In this work, we investigate a time-reversal invariant vortex in a two-dimensional helical superconductor ($d=2$). 
Assuming a finite disk geometry, we analytically solve the Bogoliubov-de-Genne (BdG) equation in the presence of a time-reversal invariant vortex with a general winding number $n$.
Although it has already been known in the case of $n=1$~\cite{Qi2009}, we find that the zero-energy Majorana Kramers pair appears at the vortex core, only when the winding number is an odd integer.
It shows that a time-reversal invariant vortex is topologically classified by $\mathbb{Z}_2$, which is consistent with the general classification theory of topological defects~\cite{Teo2010}.


More interestingly, we find that the time-reversal invariant vortex induces an emergent gravity.
By explicitly converting the BdG equation into the Dirac equation coupling to a gravitational field,
we find that the gravitational curvature is induced at the position of the vortex core, and its total flux is quantized in integer multiples of $\pi$, representing the $\mathbb{Z}_2$ topology.
Although there is no gravitational curvature except for the location of the vortex, the total curvature flux affects the energy spectrum of the edge states because of the ''gravitational" Aharonov-Bohm (AB) effect.
Indeed, when the total curvature flux is $\mathbb{Z}_2$-nontrivial, edge-localized zero modes emerge, which are topologically paired with vortex-localized zero modes.
The schematic figure of the gravitational AB effect is presented in Fig.~\ref{figure:schematic}.

Furthermore, we extend our theory to a three-dimensional helical superconductor ($d=3$) such as He-B phase. 
In this three-dimensional system, a time-reversal invariant vortex extends into a line-like object, in contrast to its point-like nature in two dimensions.
Consequently, two vortex rings can form two topologically distinct configurations: linked or unlinked [Fig.~\ref{figure:vortex_in_3D}]. 
When they are linked, the zero-energy Majorana Kramers pair emerges at each vortex, whereas it disappears when they are unlinked.~\cite{Qi2009}
We demonstrate that this phenomenon is naturally captured within our gravitational framework as a discontinuous change in the gravitational AB phase from 0 to $\pi$.

Emergent gravity in condensed matter systems has also been investigated in various contexts~\cite{Volovik1990,Unruh_PRD1995,Volovik1998,ReadGreen2000,Volovik2009,Wang2011,Parente2011,Weinfurtner_PRL2011,Ryu2012,Hughes2013,Parrikar2014,Kvorning_PRL2018,Liang2018,Golan2018,Spaansl_PRB2018,Nissinen2019,Jiang_PRL2020,Jiang_PRL2022,Tolosa_PRA2022,Chojnacki_PRB2024}, for example, in Weyl metamaterials~\cite{Weststrom2017,Guan2017,Liang2019,Nissinen2020,Jia2021} and strained graphenes~\cite{deJuan2012,deJuan2013,Arias2015,YangBo2015,Roberts2024}.
More recently, emergent gravity arising from the quantum mechanical coupling between itinerant electrons and localized spins has been proposed.~\cite{Onishi2025}

The paper is organized as follows.
In Sec.~\ref{sec:TRI vortex}, we directly solve the Bogoliubov-de-Genne equation in the presence of a time-reversal invariant vortex with arbitrary winding numbers $n$, and show that the system is described by a $\mathbb{Z}_2$ topological number.
In Sec.~\ref{sec:general gravitational holonomy}, we show that the BdG Hamiltonian with the time reversal invariant vortex can be mapped to the Dirac Hamiltonian coupling to a background gravitational field.
By using this correspondence, the $\mathbb{Z}_2$ topology of the time-reversal invariant vortex can be understood by a quantization of gravitational Aharonov-Bohm phase.
A conclusion is given in Sec.~\ref{sec:conclusion}.


\section{$\mathbb{Z}_2$ topology of time-reversal invariant vortex}\label{sec:TRI vortex}
As a starting point, we review the theoretical description of class DIII topological superconductors in Sec.~\ref{subsec:class DIII}, and introduce the time-reversal invariant vortex in Sec.~\ref{subsec:time reversal invariant vortex}. 
Then, in Sec.~\ref{subsec:helical Majorana zero mode}, we proceed to analyze the topological properties of the time-reversal invariant vortex in two-dimensional systems.
The existence of the zero-energy Majorana Kramers pair at the vortex core depends on whether the winding number $n$ is odd or even integers, resulting in the $\mathbb{Z}_2$ topology.

\begin{table}[t]
    \centering
    \begin{tabularx}{0.48\textwidth}{Y|YY}
        \hline\hline
        \textbf{} & \makecell{\textbf{Conventional} \\ \textbf{U(1) vortex}} & \makecell{\textbf{Time-reversal} \\ \textbf{invariant vortex}} 
        \\[4pt]
        \hline
        \\
        Order parameter & $e^{in\theta}\in\mathrm{U(1)}$ & $e^{in\sigma_2\theta}\in\mathrm{SO(2)}$
        \\[4pt]
        Circulating current & Electric current & Spin current
        \\[4pt]
        Quantized flux & Magnetic field & Gravitational curvature
        \\[4pt]
        Geometric phase & AB phase & Gravitational AB phase
        \\[4pt]
        \bottomrule
        \hline\hline
    \end{tabularx}
    \caption{Summary of (left) the conventional U(1) vortex and (right) the time-reversal invariant vortex in superconductors.
    In a conventional U(1) vortex, a phase gradient of the superconducting order parameter makes an effective U(1) vector potential around the vortex, and the quantized flux of the effective magnetic field is induced at the vortex core. 
    In the time-reversal invariant vortex, there emerges the spin connection $\Omega_\mu$ [Eq.~\eqref{eq:general large omega}] circulating around the vortex, and the quantized flux of the gravitational curvature $R_{12}^{12}$~[Eq.~\eqref{eq:field strength}] is induced at the vortex core.
    }
    \label{table:comparing_vortices}
\end{table}
\subsection{Class DIII topological superconductors}\label{subsec:class DIII}
We consider a $d$-dimensional time-reversal invariant topological superconductor which belongs to the class DIII in Altland Zirnbauer (AZ) symmetry classes.
Our primary focus in this paper is on the two-dimensional case ($d=2$).
The extension to the three-dimensional case ($d=3$) is addressed in Sec.~\ref{subsec:extension to 3D}.
The mean field Hamiltonian is given by $H_\mathrm{BdG}=1/2\sum_{k}\psi_k^\dagger H_\mathrm{BdG}(\vb{k})\psi_k$, where $\psi_k=(c_k,c_{-k}^\dagger)^t$ is the Nambu spinor and $H_\mathrm{BdG}(\vb{k})$ is given by
\begin{equation}
    H_\mathrm{BdG}(\vb{k})=
    \begin{pmatrix}
        \frac{\hbar^2\vb{k}^2}{2m}-\mu & \Delta_{a}^{~\mu}k_\mu\sigma^a(-i\sigma_2)\\
        (+i\sigma_2) \Delta_{a}^{~\mu}k_\mu\sigma^a& -\frac{\hbar^2\vb{k}^2}{2m}+\mu\\
    \end{pmatrix},\label{eq:H(k)}
\end{equation}
where $\mu$ is a chemical potential and $\Delta_a^{~\mu}\in\mathbb{R}$ is a superconducting pair potential.
The indices $\mu,a$ take integer values from 1 to $d$.
The system has both the particle-hole symmetry and the time-reversal symmetry,
\begin{equation}
    CH_\mathrm{BdG}(\vb{k})C^{-1}=-H_\mathrm{BdG}(-\vb{k}),~TH_\mathrm{BdG}(\vb{k})T^{-1}=H_\mathrm{BdG}(-\vb{k}),
\end{equation}
where
$C=(\sigma_1\otimes1)K$ and $T=(1\otimes -i\sigma_2)K$.
Here $K$ represents the operator that performs complex conjugation.
The preservation of time-reversal symmetry is ensured by imposing the reality condition on $\Delta_a^{~\mu}\in\mathbb{R}$.

Throughout the following discussion, we focus only on the topologically protected low-energy modes, such as vortex bound states and edge states. 
Then, at sufficiently small chemical potential, we can safely neglect the quadratic dispersion in the normal-state Hamiltonian, as it does not affect the topology of the system.
See Ref. ~\cite{ReadGreen2000,SatoFujimoto2016}
and
\footnote{
For example, in the study of topological insulators, Ref.~\cite{Aoki:2023lqp} includes the quadratic term in the computation. Then, it is shown that this quadratic term not only leaves the topological nature of the system intact but also serves to properly regularize the short-distance behavior of the edge and vortex localized modes.
}
for more details.
Thus, Eq.~\eqref{eq:H(k)} reduces to
\begin{equation}
    H_\mathrm{BdG}(\vb{k})\rightarrow
    \begin{pmatrix}
        -\mu & \Delta_{a}^{~\mu}k_\mu\sigma^a(-i\sigma_2)\\
        (+i\sigma_2) \Delta_{a}^{~\mu}k_\mu\sigma^a& +\mu\\
    \end{pmatrix}.\label{eq: simplified H(k)}
\end{equation}

A notable advantage of employing this simplified Hamiltonian is that it can be cast into the form of a Dirac Hamiltonian
\begin{equation}
    H_\mathrm{Dirac}(\vb{k})=-\gamma^0\qty(\mu-\sum_{a=1}^d\gamma^a\Tilde{k}_a).
\end{equation} 
where we define $\Tilde{k}_a\equiv\Delta_a^{~\mu}k_\mu$ and the gamma matrices are given by
\begin{equation}
    \gamma^0=\sigma_3\otimes1,~\gamma^1=i\sigma_2\otimes\sigma_3,~\gamma^2=-i\sigma_1\otimes1,~\gamma^3=-i\sigma_2\otimes\sigma_1.\label{eq:gamma matrices}
\end{equation}
They satisfy the Clliford algebra $\pb{\gamma^a}{\gamma^b}=2\eta^{ab}$, where $\eta^{ab}=\mathrm{diag}(+,-,-,-)$. 
The connection between the Dirac Hamiltonian and the BdG Hamiltonian will become particularly important in Sec.~\ref{sec:general gravitational holonomy}.



\subsection{Time-reversal invariant vortex}\label{subsec:time reversal invariant vortex}
Based on this simplified Hamiltonian~[Eq.~\eqref{eq: simplified H(k)}], we examine a 2D finite disk-shaped system of radius $r=R$, that has a time-reversal invariant vortex at $r=0$.
See Fig.~\ref{figure:schematic}~(a).
This system can be modeled by allowing the chemical potential to vary with position, i.e., by substituting $\mu\rightarrow\mu(\vb{r})$
with
\begin{equation}
    \mu(\vb{r})\equiv
    \begin{cases}
    +\mu_0>0 & \text{if $0<r<R$,} \\
    -\mu_0<0 & \text{if $r>R$.}
    \end{cases}
\end{equation}
For the off-diagonal pairing term, we replace the momentum with its operator form,
$k_\mu\rightarrow-i\partial_\mu$,
and allow the pairing potential to vary spatially,
$\Delta_a^{~\mu}\rightarrow\Delta_a^{~\mu}(\vb{r})$.
Explicit expression for $\Delta_a^{~\mu}(\vb{r})$ describing the time-reversal invariant vortex will be given in the next paragraph.~(See Eq.~\eqref{eq:TRI_vortex}.)
We then take the anticommutator between these two quantities, and obtain
\begin{equation}
    \Delta(\vb{r})\equiv\frac{1}{2}\acomm{\Delta_a^{~\mu}(\vb{r})}{-i\partial_\mu}~\sigma^a\qty(-i\sigma_2).\label{eq:anticomm}
\end{equation}
With these steps, the simplified BdG Hamiltonian [Eq.~\eqref{eq: simplified H(k)}] can be expressed in its final form as
\begin{equation}
    H_\mathrm{BdG}(\vb{r})=\begin{pmatrix}
        -\mu(\vb{r}) & \Delta(\vb{r})\\
        \Delta^\dagger(\vb{r}) & \mu(\vb{r})\label{eq:BdG Hamiltonian}
    \end{pmatrix}.
\end{equation}

The BdG Hamiltonian in real space is guaranteed to satisfy particle–hole symmetry and time-reversal symmetry,
\begin{equation}
    CH_\mathrm{BdG}(\vb{r})C^{-1}=-H_\mathrm{BdG}(\vb{r}),~TH_\mathrm{BdG}(\vb{r})T^{-1}=H_\mathrm{BdG}(\vb{r}),
\end{equation}
provided that the superconducting order parameter is real, $\Delta_a^{~\mu}(\vb{r})\in\mathbb{R}$.
In the case of a conventional vortex $\Delta_{a}^{~\mu}(\vb{r})=\Delta_0 \bm{1}e^{in\theta}$, it winds the U(1) phase of the order parameter, giving it a complex value.
Hence, it inevitably breaks the time-reversal symmetry.
In contrast, we consider a time-reversal invariant vortex~\cite{Qi2009} that winds the spin degrees of freedom instead of the U(1) phase, giving
\begin{align}
    \Delta_{a}^{~\mu}(\vb{r})&=\Delta_0e^{i\sigma_2n\theta}\notag\\
    &=\Delta_0\mqty(
    \cos n\theta & \sin n\theta\\
    -\sin n\theta & \cos n\theta
    ).
    \label{eq:TRI_vortex}
\end{align}
A striking feature of the time-reversal invariant vortex is that the order parameter remains real throughout the system.
Table~ \ref{table:comparing_vortices} summarizes and compares the physical properties of the conventional U(1) vortex and the time-reversal invariant vortex.

It should be noted that a two-dimensional helical superconductor can be described as a system in which the spin-up component forms a $p_x-ip_y$ chiral superconductor, while the spin-down component forms a $p_x+ip_y$ chiral superconductor.~\cite{Qi2009}
Based on this picture, the Hamiltonian [Eq.~\eqref{eq:BdG Hamiltonian}] can be transformed into a block-diagonal form in the spin basis, where the spin up $p_x-ip_y$ chiral superconductor has a vortex $\Delta_0e^{in\theta}$, while the spin-down $p_x+ip_y$ chiral superconductor has a vortex with opposite winding number, $\Delta_0e^{-in\theta}$.
Since the electric currents around the vortex in the spin-up and spin-down sectors flow in opposite directions, the net electric current vanishes, while a spin current remains. 
Therefore, the time-reversal-invariant vortex can be interpreted as a vortex of spin current.
However, such block diagonalization by spin basis is only possible in the two-dimensional case; in three dimensions, it becomes impossible due to the presence of the $p_z$-component.
The case of the three dimensional system is discussed in Sec.~\ref{subsec:extension to 3D}.

\subsection{Majorana Kramers pairs}\label{subsec:helical Majorana zero mode}
We concentrate on a vortex bound state localized at a core of a time-reversal invariant vortex.
In Ref.~\cite{Qi2009}, the time-reversal invariant vortex with $n=1$ [Eq.~\eqref{eq:TRI_vortex}] is studied, and the existence of the zero-energy Majorana Kramers pair localized at the vortex core is demonstrated.
This zero-energy Majorana Kramers pair is made up of spin-up and spin-down Majorana fermions in $p_x\pm ip_y$ chiral superconductors.~\cite{SatoFujimoto2016,Gurarie2007,Tewari2007}

Here, we extend the argument to the general winding number $n$.
We explicitly solve the BdG equation $H_\mathrm{BdG}(\vb{r})\psi=E\psi$, and find the energy spectrum of the vortex-bound state and the edge state
as illustrated in Fig.~\ref{figure:spectrum}.
Detailed calculations are presented in the Appendix~\ref{subsec:BdG equation}.
Defining an effective angular momentum around the $z-$axis as
\begin{equation}
    J=-i\pdv{\theta}-\frac{n-1}{2}\sigma_3\otimes\sigma_3,\label{eq:angular momentum operator}
\end{equation}
it commutes with the BdG Hamiltonian.
Consequently, the energy spectrum can be expressed as a function of its eigenvalue $j$.
Since the term $-(\sigma_3\otimes\sigma_3)/2=\diag(-1/2,1/2,1/2,-1/2)$ in Eq.~\eqref{eq:angular momentum operator} corresponds to the spin component of the Nambu spinor $\psi=(c_\uparrow,c_\downarrow,c^\dagger_\uparrow,c^\dagger_\downarrow)$, it can interpreted as a Berry phase contribution arising from the spin current around $z$-axis.

The low-energy spectrum 
differs qualitatively depending on whether the winding number is odd [Fig.~\ref{figure:spectrum}(a)] or even [Fig.~\ref{figure:spectrum}(b)].
More precisely, the zero-energy Majorana Kramers pair simultaneously appears at both the vortex core and the edge only when the winding number is odd ($n=2k+1$), while there is no such zero mode when the winding number is even ($n=2k$), representing the $\mathbb{Z}_2$ topology.

In general, Majorana fermions must appear in pairs in finite systems, combining to form a complex fermion.
Such a hybridization between localized Majorana zero modes in this system is explicitly demonstrated in the Appendix~\ref{subsec:zero-mode mixing}.

This $\mathbb{Z}_2$ topological characterization of a time-reversal invariant vortex is consistent with the general classification theory of topological defects~\cite{Teo2010}:
a zero-dimensional point defect in two-dimensional class $\mathrm{DIII}$ topological superconductors possess a $\mathbb{Z}_2$ topological number.
Furthermore, a one-dimensional line defect in three-dimensional class $\mathrm{DIII}$ topological superconductors also has a $\mathbb{Z}_2$ topological number.
This point will be discussed in Sec.~\ref{subsec:extension to 3D}.



\begin{figure}[t]
    \centering
    \includegraphics{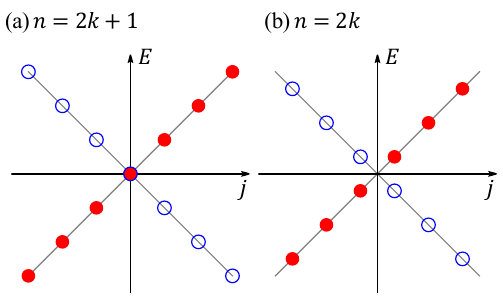}
    \caption{Schematic figure of the energy spectrum of the edge states [Eq.~\eqref{eq:energy spectrum of edge state}] and vortex bound states [Eq.~\eqref{eq:energy spectrum of vortex bound state}] when the winding number of the time reversal invariant vortex is (a) odd integer and (b) even integer.
    The red (blue) circle represents the energy eigenvalue corresponding to the spin-up (spin-down) state for edge states and spin-down (spin-up) state for vortex bound states.
    Here $j$ represents the eignevalue of the effective angular momentum $J$.
    }
    \label{figure:spectrum}
\end{figure}

\section{Mapping to the theory of gravity
}\label{sec:general gravitational holonomy}

In this section, we show that a time-reversal invariant vortex induces an effective gravitational field in the system, whose topology is characterized by a gravitational Aharonov–Bohm (AB) phase quantized in integer multiples of $\pi$.
In Sec.~\ref{subsec:general theory},
we present a general formulation of the Dirac Hamiltonian in 
$d$-dimensional curved spacetime, along with essential geometrical concepts such as the vielbein, spin connection, and gravitational curvature tensor.
In Sec.~\ref{subsec:BdG to curved Dirac}, we show that our BdG Hamiltonian with a time-reversal invariant vortex can be identified as a Dirac Hamiltonian in a nontrivial gravitational background.
In Sec.~\ref{subsec:z2 gravitational berry phase}, we find that a topological quantization of the gravitational AB phase characterizes the $\mathbb{Z}_2$ topology of the time-reversal invariant vortex.
In Sec.~\ref{subsec:extension to 3D}, the extension to a three-dimensional system is discussed.

\subsection{General theory of Dirac fermion in curved spacetime}\label{subsec:general theory}
In general relativity, the gravitational field is described by the spacetime metric $g_{\mu\nu}$ and the vielbein $e^a_{~\mu}$, 
which are related by $g_{\mu\nu}=e^a_{~\mu}e^b_{~\nu}\eta_{ab}$, where $\eta_{ab}=\mathrm{diag}(+1,-1,-1, \cdots)$ 
is the metric in the local Lorentz frame.
Here, we consider a general $d+1$-dimensional theory.


The electrons in a curved spacetime follow the Dirac equation
\begin{equation}
    \qty(i\sum_{a=0}^d\sum_{\mu=0}^d\gamma^a e_a^{~\mu} D_\mu+m)\psi=0,\label{eq: curved Dirac}
\end{equation}
where $m$ represents a mass of the electron,
$\gamma_a$'s are Dirac matrices satisfying
$\{\gamma_a,\gamma_b\}=2\eta_{ab}$.
See Eq.~\eqref{eq:gamma matrices} 
for their explicit forms.
The covariant derivative is 
\begin{equation}
    D_\mu=\partial_\mu+\Omega_\mu,
\end{equation}
where the spin connection $\Omega_\mu$ describes the coupling
of the fermion to gravity.

In general relativity, the spin connection as well as the Clistoffel symbol is not an independent quantity but a function of vielbein and metric. 
From the metricity condition and the equivalence principle, the Clistoffel symbol is uniquely given by
\begin{equation}
\Gamma_{\mu\nu}^\kappa=\sum_{\lambda=0}^d\frac{1}{2}g^{\kappa\lambda}(\partial_\mu g_{\lambda\nu}+\partial_\nu g_{\lambda\mu}-\partial_\lambda g_{\mu\nu}),\label{eq:Christoffel symbol}
\end{equation}
and the spin connection is given by
\begin{equation}
    \Omega_\mu=\sum_{a,b=0}^d\frac{1}{2}\omega^{ab}_\mu\Sigma_{ab},\label{eq:general large omega}
\end{equation}
where
\begin{equation}
    \omega^{ab}_\mu=\sum_{\nu=0}^d e^a_{~\nu}(\partial_\mu e^{b\nu}+\Gamma_{\mu\lambda}^\nu e^{b\lambda}) 
    \label{eq:general small omega}
\end{equation}
is determined by the so-called vielbein postulate, and
\begin{equation}
    \Sigma_{ab}=\frac{1}{4}[\gamma_a,\gamma_b]\label{eq:definition of Sigma_ab}
\end{equation} 
is the local Lorentz generetor (i.e. the spinor representation of SO$(1,d)$ Lie algebra).

Note that the field strength $R_{\mu\nu}^{ab}$ of the spin connection is 
\begin{equation}
\sum_{a,b=0}^d R_{\mu\nu}^{ab}\Sigma_{ab}=\partial_\mu\Omega_\nu-\partial_\nu\Omega_\mu+[\Omega_\mu,\Omega_\nu],\label{eq:field strength}
\end{equation}
which is related to the Riemann curvature tensor $R^\rho_{\sigma \mu\nu}$ by
\begin{equation}
R^\rho_{\sigma \mu\nu}=\sum_{a,b=0}^d e^{~\rho}_a e_{b\sigma}R_{\mu\nu}^{ab}.
\end{equation}

When the system is static, we can define the Dirac Hamiltonian
\begin{equation}
    H=-\gamma^0\qty(m+\sum_{a,\mu=1}^{d}i\gamma^ae_a^{~\mu}D_\mu),\label{eq:Dirac Hamiltonian in curved space}
\end{equation}
so that the Dirac equation Eq.~\eqref{eq: curved Dirac} can be converted to the conventional Schrodinger equation $i\partial_t\psi=H\psi$.

\subsection{From BdG Hamiltonian to Dirac Hamiltonian in curved space}\label{subsec:BdG to curved Dirac}

We go back to the original static $2+1$-dimensional system and the roman and greek indices below take $1$ or $2$ only.
We show that the BdG Hamiltonian in Eq.~\eqref{eq:BdG Hamiltonian} with a time-reversal invariant vortex can be rewritten as the Dirac Hamiltonian in a curved space [Eq.~\eqref{eq:Dirac Hamiltonian in curved space}] in a proper way.
In the following, we put $\Delta_0=1$ for simplicity.

The off-diagonal term $\Delta(\vb{r})$ in $H_\mathrm{BdG}(\vb{r})$ is given by Eq.~\eqref{eq:anticomm}.
By calculating the anticommutator, it can be decomposed into two components,
\begin{equation}
    \Delta(\vb{r})=\frac{1}{2}\acomm{\Delta_a^{~\mu}(\vb{r})}{-i\partial_\mu}~\sigma^a\qty(-i\sigma_2)=A(\vb{r})+B(\vb{r}),
\end{equation}
where
\begin{align}
    A(\vb{r})=&\Delta_a^{~\mu}\sigma^a(-i\sigma^2)(-i\partial_\mu)\notag\\
    =&(\cos n\theta~\sigma^3+i\sin n\theta)(-i\partial_1)\notag\\
    &+(\sin n\theta~\sigma^3-i\cos n\theta)(-i\partial_2)
\end{align}
and
\begin{align}
    B(\vb{r})=&\frac{1}{2}(-i\partial_\mu\Delta_a^{~\mu})\sigma^a(-i\sigma^2)\notag\\
    =&-i\frac{n\partial_1\theta}{2}(-\sin n\theta~\sigma^3+i\cos n\theta)\notag\\
    &-i\frac{n\partial_2\theta}{2}(\cos n\theta~\sigma^3+i\sin n\theta).
\end{align}
From $\Delta^\dagger=-\Delta^*$, we have the following equalities,
\begin{align}
    \begin{pmatrix}
        -\mu(\vb{r}) & 0\\
        0 & \mu(\vb{r})
    \end{pmatrix}
    &=-\mu(\vb{r})\gamma^0,\\
    \begin{pmatrix}
        0 & A(\vb{r})\\
        -A^*(\vb{r}) & 0
    \end{pmatrix}
    &=\gamma^0\gamma^a\Delta_a^{~\mu}(-i\partial_\mu),\\
    \begin{pmatrix}
        0 & B(\vb{r})\\
        -B^*(\vb{r}) & 0
    \end{pmatrix}
    &=\gamma^0\gamma^a\Delta_a^{~\mu}\qty(-in\partial_\mu\theta\Sigma_{12}),
\end{align}
where the definition of the gamma matrices is presented in \eqref{eq:gamma matrices}.
Thus, the BdG Hamiltonian becomes
\begin{equation}
    H_\mathrm{BdG}=-\gamma^0\qty(\mu+\sum_{a,\mu=1,2}i\gamma^a\Delta_a^{~\mu}\qty(\partial_\mu+\Omega'_\mu)),\label{eq:BdG to curved-Dirac}
\end{equation}
where
\begin{equation}
    \Omega'_\mu
    =n\partial_\mu\theta\Sigma_{12}.\label{eq:spin connection(BdG)}
\end{equation}

Now let us compare the Hamiltonian with the Dirac Hamiltonian in curved space [Eq.~\eqref{eq:Dirac Hamiltonian in curved space}].
It is natural to assume that $\Delta_a^{~\mu}$ 
corresponds to the vielbein $e_a^{~\mu}$.
However, in order to establish the exact correspondence with the theory of gravity, it is necessary to show that the obtained connection
$\Omega'_\mu$ coincides with the spin connection
$\Omega_\mu$, which is uniquely determined
by the vielbein $e_a^{~\mu}=\Delta_a^{~\mu}$ using Eq.~\eqref{eq:general large omega}
and Eq.~\eqref{eq:general small omega}.

To this end, we first calculate the Christoffel symbols.
The spacial components of the metric is given by
\begin{align}
    g^{\mu\nu}&=\Delta^{~\mu}_a\Delta^{~\nu}_b\eta^{ab}\notag\\
    &=-\mqty(
    \cos n\theta & \sin n\theta\\
    -\sin n\theta & \cos n\theta
    )^t
    \mqty(
    \cos n\theta & \sin n\theta\\
    -\sin n\theta & \cos n\theta
    )\notag\\
    &=-
    \mqty(
    1 & 0\\
    0 & 1
    ).
\end{align}
By incorporating the time component, the full spacetime metric can be expressed as
\begin{align}
    ds^2&=g_{\mu\nu}dx^\mu dx^\nu\notag\\
    &=dt^2-dx^2-dy^2.
\end{align}
Therefore, the spacetime is flat and coincides with Minkowski spacetime everywhere except at the origin, where a singularity is present.
As a result, the Christoffel symbols [Eq.~\eqref{eq:Christoffel symbol}] vanish identically,
\begin{equation}
    \Gamma^\kappa_{\mu\nu}=0.
\end{equation}
It is (trivially) consistent with the equivalence principle.
It should be noted that the winding number of the vielbein $n$ (i.e. topology of gravity) is not reflected in the space-time metic and the Christoffel symbol.

Next we explicitly confirm that $\Omega'_\mu$ given in Eq.~\eqref{eq:spin connection(BdG)} coincides with the spin connection.
Substituting $\Gamma^\mu_{\nu\rho}=0$ and $e_a^{~\mu}=\Delta_a^{~\mu}$ in Eq.~\eqref{eq:general small omega},
we have
\begin{align}
    \omega_\mu^{12}&=
    \Delta^1_{~\nu}\partial_\mu \Delta^{2\nu}\notag\\
    &=
    \begin{pmatrix}
        \cos n\theta & \sin n\theta
    \end{pmatrix}
    \partial_\mu
    \begin{pmatrix}
        \sin n\theta\\
        -\cos n\theta
    \end{pmatrix}
    \notag\\
    &=n\partial_\mu\theta.\label{eq:omega_mu^12}
\end{align}
Thus, $\Omega'_\mu$ given in Eq.~\eqref{eq:spin connection(BdG)} can be identified as the 
spin connection $\Omega_\mu$ with respect to the SO(2) part of the local Lorentz symmetry and the BdG
Hamiltonian can be interpreted as the Dirac Hamiltonian with a gravitational background.

It is also interesting to note that in the polar coordinate, the connection which can be read from Eq.~(\ref{eq:Ht}) is proportional to $n-1$ rather than $n$.
This reflects that the edge of the circle with radius $R$ itself is curved.
This additional gravitational effect is proportional to $-1$. See \cite{Aoki:2022aez, Aoki:2022cwg} for the details of the induced spin connection due to the curved surface.

\subsection{Quantization of gravitational Aharonov-Bohm phase and $\mathbb{Z}_2$  topology}\label{subsec:z2 gravitational berry phase}

Using the formal correspondence established in Sec.~\ref{subsec:BdG to curved Dirac}, we evaluate the gravitational $\mathbb{Z}_2$ topology of the time-reversal invariant vortex.

Let us compute the gravitational curvature tensor. 
Since the $\Sigma_{12}$ component is the only nonzero contribution to the spin connection,  
the gravitational curvature tensor [Eq.~\eqref{eq:field strength}] is computed as 
\begin{align}
    R_{12}^{12}&=\frac{1}{2!}(\partial_1\omega_2^{12}-\partial_2\omega_1^{12})\notag\\
    &=n\pi\delta(\bm{r}),\label{eq: R_12^12}
\end{align}
where we have used $[\Omega_\mu,\Omega_\nu]=0$, and $\delta(\vb{r})$ is the Dirac delta function.
There is a $n\pi$-flux of the curvature tensor at the vortex core, although it is zero everywhere except at this point.
The behavior corresponds to the fact that the metric is singular at the vortex core and is locally flat at all other locations.

Nevertheless $R_{12}^{12}=0$ except at the origin, the fermion field at $r\neq 0$ receives a nontrivial gravitational contribution, that is nothing but the gravitational AB effect.
One can show this by integrating the spin connection $\Omega_\mu$ along a circle with radius $R$, giving
\begin{equation}
    \oint_{r=R} \Omega_\mu dx^\mu
    =\int_{r<R} d^2x R^{ab}_{12}\Sigma_{ab}  
    =-i n\pi(\sigma_3\otimes\sigma_3),
\end{equation}
where we used Stokes theorem and $\Sigma_{12}=-i/2(\sigma_3\otimes\sigma_3)$.
The obtained gravitational AB phase becomes nontrivial only when $n$ is an odd integer, representing the $\mathbb{Z}_2$ topology.


With this gravitational version of the Aharonov-Bohm effect, the spin connection at $r=R$ does affect the Dirac operator spectrum.
As we have seen in the edge state spectrum localized at $r=R$ in Sec.~\ref{sec:TRI vortex}, the spin connection in the effective angular momentum $J$ [Eq.~\ref{eq:angular momentum operator}] is proportional to $n-1$ and the value modulo 2 determines if the Dirac operator can have zero modes or not.
It is interesting to note that even when $n=0$, we have a nontrivial gravitational effect, 
which is induced \cite{Aoki:2022cwg, Aoki:2022aez} by the curved edge with radius $r=R$ embedded into the $\mathbb{R}^2$ space.

As discussed in \cite{Aoki:2022aez, Aoki:2023lqp,Aoki:2024bwx}, the vortex-localized modes can be identified as another edge states sitting on a domain wall with radius $r=r_0$, which is created near the vortex.
These zero-modes localized at a vortex core appear only when $n$ is odd and always make a pair with one of the edge zero modes at $r=R$.

Thus, the origin of the $\mathbb{Z}_2$ structure of time reversal invariant vortex can be attributed to the gravitational Aharonov-Bohm phase, originating from the $n\pi$-flux of the gravitational curvature tensor.
When $n$ is an odd integer, in a finite disc geometry, a Majorana Kramers pair of zero-modes is isolated on the edge, and another Majorana Kramers pair of zero-modes appears at the location of the vortex, and they maximally hybridize to form a non-local complex fermion.

\begin{figure}[t]
    \centering
    \includegraphics[width=1\linewidth]{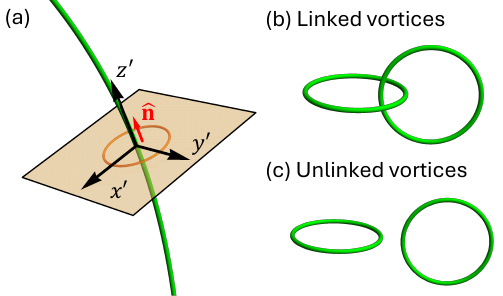}
    \caption{
    A time-reversal invariant vortex in a three-dimensional helical superconductor.
    (a) The one-dimensional vortex structure $\vb{r}_0(z')$ is depicted by the green curve.
    For every fixed $z'$, one can define a local coordinate system whose $x'y'$-plane is perpendicular to the local orientation $\hat{\vb{n}}$ of the vortex.
    (b,c) In a closed three-dimensional system, 
    the topology of a vortex pair configuration is determined by whether the two vorteex rings are linked or unlinked. 
    When a linking occurs, the system undergoes a topological phase transition, manifested as a discontinuous jump of the gravitational AB phase along one vortex from $0$ to $\pi$.
    }
    \label{figure:vortex_in_3D}
\end{figure}
\subsection{Extension to three-dimensional systems}\label{subsec:extension to 3D}
In the preceding section, by using the mapping to the theory of gravity, we elucidated the $\mathbb{Z}_2$ topology of a time-reversal invariant vortex in two-dimensional helical superconductors, where it is realized as a zero-dimensional point defect.
\\\indent
In the present section, we generalize this framework to three dimensions, wherein the time-reversal-invariant vortex corresponds to a one-dimensional line defect.
According to the general classification theory of topological defects~\cite{Teo2010}, a one-dimensional line defect in three-dimensional class DIII topological superconductors is also characterized by $\mathbb{Z}_2$. 
This distinguishes between two cases: the presence or absence of a one-dimensional helical Majorana zero mode localized along the line defect.
\\\indent
The key difference from the two dimensional case is that the vortices in three dimensions are line-like objects, which can extend 
along an arbitrary curve, represented by $\vb{r}_0(z')$, with $z'$ denoting a real-valued parameter.
The schematic figure is presented in Fig.~\ref{figure:vortex_in_3D}~(a).
For a fixed value of $z'$, we introduce a local coordinate system $\vb{r}'=(x',y',z')$, where the $z'$-axis is aligned with the tangent vector
\begin{equation}
    \hat{\vb{n}}=\pdv{\vb{r}_0(z')}{z'}=(n_x,n_y,n_z),
\end{equation}
where $|\hat{\vb{n}}|=1$.
$\theta'$ denotes the angular coordinate that parametrizes the circular path around the $z'$-axis in the $x'y'$-plane.
\\\indent
With this setup, the order parameter of such a time-reversal invariant vortex with a winding number $n$ is given by
\begin{equation}
    \Delta_a^{~\mu}(\vb{r})
    =R(z')\,
    \mqty(
    \cos n\theta' & \sin n\theta' & 0\\
    -\sin n\theta' & \cos n\theta' &0\\
    0 & 0 & 1
    )
    \,R^{-1}(z'),\label{eq: vielbein in 3D}
\end{equation}
where $R(z')\in\mathrm{SO(3)}$ is a rotation matrix which transforms the $x'y'z'$-coordinate system into the $xyz$-coordinate system.
In general, it can be constructed using the method of Euler angles.
Eq.~\eqref{eq: vielbein in 3D} represent natural generalizations of Eq.~\eqref{eq:TRI_vortex}, with the underlying matrix group being SO(3) rather than SO(2).
\\ \indent
In a manner completely analogous to the two-dimensional case, the BdG Hamiltonian with the time-reversal invariant vortex can be transformed into a Dirac Hamiltonian in curved space-time by identifying the order parameter $\Delta_a^{~\mu}$ with the vielbein $e_a^{~\mu}$.
Then, the spin connection [Eq.~\eqref{eq:general large omega}] is given by
\begin{equation}
    \Omega_\mu=\omega_\mu^{'12}\Sigma_{12}'=-in\partial_\mu'\theta'(\hat{\vb{n}}\cdot\vb{\Sigma}),\label{eq:spin connection of 1D vortex}
\end{equation}
where we set
\begin{equation}
    \vb{\Sigma}
    =i(\Sigma_{23},\Sigma_{31},\Sigma_{12}),    
\end{equation}
where $\Sigma_{23}=-i/2(\sigma_3\otimes\sigma_1),~\Sigma_{31}=-i/2(1\otimes\sigma_2),~\Sigma_{12}=-i/2(\sigma_3\otimes\sigma_3)$.
They form a spinor representaion of the SO(3) Lie algebra, satisfying the relation $[\Sigma_a,\Sigma_b]=i\epsilon_{abc}\Sigma_c$.
As in the case of the two-dimensional system, there emerges the spin connection circulating around the time-reversal invariant vortex. 
However, the distinction lies in the fact that the matrix-valued term $\hat{\vb{n}}\cdot\vb{\Sigma}$ in Eq.~\eqref{eq:spin connection of 1D vortex} is determined by the local orientation $\hat{\vb{n}}$ of the one-dimensional vortex.
\\ \indent
Subsequently, the gravitational curvature tensor is calculated as
\begin{equation}
    R_{12}^{12'}=n\pi\delta(\vb{r}-\vb{r}_0(z'))
\end{equation}
in a same way to Eq.~\eqref{eq: R_12^12}.
The $n\pi-$flux of the gravitational curvature tensor is induced at the position of the vortex.
\\\indent
Based on the discussions presented so far, we can now get the topological invariant characterizing the time-reversal invariant vortex.
This is nothing but the gravitational AB phase: the contour integral of the spin connection around the vortex,
\begin{equation}
    \exp\qty(\oint \Omega_\mu dx^\mu)
    =(-1)^n
\end{equation}
We see that the gravitational AB phase becomes non-trivial only when the winding number $n$ is an odd integer, representing $\mathbb{Z}_2$ topology.
These results show that our mapping to the theory of gravity  is also valid in three-dimensional systems, and offers an intuitive way to understand the topological properties of the time-reversal invariant vortex.
\\\indent
Moreover, in a closed three-dimensional system, a pair of the ring-shaped time-reversal invariant vortex is possible. 
Two such vortex rings with odd winding numbers may form either a linked or an unlinked configuration. 
See Fig.~\ref{figure:vortex_in_3D}~(b) and (c).
They are topologically distinct from each other and a zero-energy Majorana Kramers pair emerges only when the vortex rings are linked.~\cite{Qi2009,Lopes2017}
Our gravitational picture naturally describes the $\mathbb{Z}_2$ topology as follows.
The gravitational Aharonov–Bohm phase associated with a given vortex ring undergoes a discontinuous transition from $0$ to $\pi$ upon the linking of two initially unlinked vortex rings, reflecting the topological phase transition.
\\\indent
Note that in this three-dimensional case, due to the $z$-dependence, the system is no longer equivalent to a direct product of $(p_x\pm ip_y)$-chiral superconductors, as mentioned in the last paragraph of Sec.~\ref{subsec:time reversal invariant vortex}.

\section{Conclusion}\label{sec:conclusion}
We have studied a time-reversal invariant vortex (i.e. spin vortex) in a class DIII topological superconductor by proposing the mapping to the theory of gravity.
Identifying a superconducting order parameter as a veilbein, we map the BdG Hamiltonian into the Dirac Hamiltonian in curved spacetime.
In this context, at the core of the time-reversal invariant vortex of winding number $n$, the $n\pi$-flux of the gravitational curvature is induced.
Accordingly, the resulting gravitational AB phase becomes 0 or $\pi$, reflecting the $\mathbb{Z}_2$ topological structure.
We also demonstrated this gravitational AB phase can be utilized to characterize a vortex linking process in a three-dimensional system.


As a future direction, it would be intriguing to investigate whether emergent gravity associated with a spin texture can arise in other types of superconductors.
\footnote{
Our analysis can, in principle, be extended to multi-band superconductors; however, its direct application to iron-based superconductors is not straightforward. 
This limitation arises because our approach relies on the mapping to the Dirac Hamiltonian that is linear in the wave vector $k$. Consequently, the Cooper pairing in the superconductors must be of the p-wave type.
}


\section*{Acknowledgement}
We acknowledge fruitful discussions with Takuto Kawakami and Mikito Koshino.
This work was supported in part by JSPS KAKENHI Grants No. JP23KJ1518, No. JP24KJ0157, No. JP23K22490 and No. JP25K07283, Japan.
\appendix
\section{Analytical solutions of BdG equation}\label{subsec:BdG equation}
Assuming the finite disk geometry with radius $r=R$, we directly solve the BdG equation
\begin{equation}
    H_\mathrm{BdG}(\vb{r})\psi=E\psi\label{eq:BdG}
\end{equation}
in the presence of the time-reversal invariant vortex at $r=0$ with general winding number $n$.
Since the effective angular momentum $J$ [Eq.~\eqref{eq:angular momentum operator}] commutes with $H_\mathrm{BdG}(\vb{r})$, the eigenvalue $j$ of $J$ becomes a good quantum number,
\begin{equation}
    J\psi=j\psi.\label{eq:J=j}
\end{equation}

It is convenient to separate  the BdG Hamiltonian [Eq.~\eqref{eq:BdG Hamiltonian}] into radial and angular parts, giving $H_\mathrm{BdG}(\vb{r})=H_r(\vb{r})+H_\theta(\vb{r})$, where
\begin{align}
    H_r(\vb{r})&=(\sigma_3\otimes1)\Biggl[-\mu(r)+\Delta_0\Gamma \qty(\pdv{r}+\frac{1}{2r})\Biggr],\label{eq:Hn}\\
    H_\theta(\vb{r})&=(1\otimes\sigma_3)\Delta_0\Gamma \qty(-\frac{i}{r}\pdv{\theta}-\frac{n-1}{2r}\sigma_3\otimes\sigma_3),
    \label{eq:Ht}
\end{align}
where $\Gamma=(\sigma_1\otimes1)\exp[i(\pi/2-(n-1)\theta)(\sigma_3\otimes\sigma_3)]$
is the chirality operator.
As shown below, the edge and vortex Majorana zero modes have chiralities +1 and -1, respectively.


\subsection{Edge states}

The exact solutions of Eq.~\eqref{eq:BdG} that grow exponentially for large $r$ are given as follows.
The spin-up states are expressed as
\begin{equation}
    \psi_{+\uparrow j}=
    \begin{pmatrix}
        \sqrt{1-\frac{E}{\mu_0}}~I_{j-\frac{1}{2}}(\Bar{r})e^{-i\frac{\pi}{4}+i(j+\frac{n-1}{2})\theta}\\
        0\\
        \sqrt{1+\frac{E}{\mu_0}}~I_{j+\frac{1}{2}}(\Bar{r})e^{i\frac{\pi}{4}+i(j-\frac{n-1}{2})\theta}\\
        0
    \end{pmatrix},
\end{equation}  
and the spin-down states are expressed as
\begin{equation}
    \psi_{+\downarrow j}=
    \begin{pmatrix}
        0\\
        \sqrt{1-\frac{E}{\mu_0}}~I_{j+\frac{1}{2}}(\Bar{r})e^{i\frac{\pi}{4}+i(j-\frac{n-1}{2})\theta}\\
        0\\
        \sqrt{1+\frac{E}{\mu_0}}~I_{j-\frac{1}{2}}(\Bar{r})e^{-i\frac{\pi}{4}+i(j+\frac{n-1}{2})\theta}
    \end{pmatrix},
\end{equation} 
where we define a dimensionless quantity,
\begin{equation}
    \Bar{r}\equiv\frac{\mu_0r}{\Delta_0}\sqrt{1-\qty(\frac{E}{\mu_0})^2}.
\end{equation}

Here $I_\nu(x)~(=I_{-\nu}(x))$ is the modified Bessel function of the first kind.
These functions of neighboring orders are connected by the raising and lowering operators, as given in
\begin{align}
    I_{\nu-1}(x)&=\qty(\pdv{x}+\frac{\nu}{x})I_\nu(x),\\
    I_{\nu+1}(x)&=\qty(\pdv{x}-\frac{\nu}{x})I_\nu(x).
\end{align}
The asymptotic behaviors are given by
\begin{equation}
    I_\nu(x)\approx\frac{e^{x}}{\sqrt{x}}~(x\gg1).
\end{equation}

As required by the single-valuedness of the wave function, $j$ takes integers or half-integers depending on whether the winding number $n$ is odd or even.
Hence, the $j=0$ zero-energy state appears only for odd $n$.
The subscript ''+" in the wave function $\psi_{+\alpha j}~(\alpha=\uparrow\downarrow)$ originates from the fact that the $j=0$ zero-energy state has chirality +1, satisfying $\Gamma\psi_{+\alpha 0}=+\psi_{+\alpha 0}$.

We focus on the low-energy edge states satisfying $E\ll\mu_0$.
Thus, we employ a low-energy effective Hamiltonian for edge states, obtained by setting $r=R$ in the angular part [Eq.~\eqref{eq:Ht}] of the original Hamiltonian. 
In this limit, the expressions for the energy spectrum $E_{+\alpha j}~(\alpha=\uparrow\downarrow)$
are approximated to be
\begin{equation}
    E_{+\uparrow j}\approx\frac{\Delta_0}{R}j,
    \quad 
    E_{+\downarrow j}\approx-\frac{\Delta_0}{R}j.
    \label{eq:energy spectrum of edge state}
\end{equation}
The schematic pictures of the energy spectrum are illustrated in Fig.~\ref{figure:spectrum}.
The spin-up edge states are given by
\begin{equation}
    \psi_{+\uparrow j}\approx f(r)
    \begin{pmatrix}
        e^{-i\frac{\pi}{4}+i(j+\frac{n-1}{2})\theta}\\
        0\\
        e^{i\frac{\pi}{4}+i(j-\frac{n-1}{2})\theta}\\
        0
    \end{pmatrix},
\end{equation}  
and the spin-down edge states are given by
\begin{equation}
    \psi_{+\downarrow j}\approx f(r)
    \begin{pmatrix}
        0\\
        e^{i\frac{\pi}{4}+i(j-\frac{n-1}{2})\theta}\\
        0\\
        e^{-i\frac{\pi}{4}+i(j+\frac{n-1}{2})\theta}
    \end{pmatrix},
\end{equation}  
where
\begin{equation}
    f(r)=\frac{1}{\sqrt{r}}\exp\qty[-\frac{\mu_0}{\Delta_0}(R-r)].    
\end{equation}
In this approximation, all of these states have a definite chirality $\Gamma\psi_{+\alpha j}=+\psi_{+\alpha j}$, not only for the $j=0$ state but also for all $j$.

Degenerate states with opposite spins form a Kramers pair, that is related by time reversal transformation as
\begin{equation}
    T\psi_{+\uparrow ,j}=\psi_{+\downarrow,-j},~T\psi_{+\downarrow ,-j}=-\psi_{+\uparrow,j}.
\end{equation}
The eigenstates having the same spin but having the opposite sign of $j$ are a particel-hole pair, that is related via
particle-hole transformation as 
\begin{equation}
    C\psi_{+\uparrow ,j}=\psi_{+\uparrow,-j},~C\psi_{+\downarrow ,j}=\psi_{+\downarrow,-j}.
\end{equation}
The important observation is that, when the winding number $n$ is an odd integer, $j=0$ is possible, enabling the existence of the zero-energy Majorana Kramers pair,
\begin{align}
    T\psi_{+\uparrow ,0}&=\psi_{+\downarrow,0},~T\psi_{+\downarrow ,0}=-\psi_{+\uparrow,0},\\
    C\psi_{+\uparrow ,0}&=\psi_{+\uparrow,0},~C\psi_{+\downarrow ,0}=\psi_{+\downarrow,0}.
\end{align}
Being a Kramers pair, it is robust under any time reversal invariant perturbation.

\subsection{Vortex bound states}
The exact solutions of Eq.~\eqref{eq:BdG} that decay exponentially for large $r$ are given as follows.
The spin-up states are expressed as
\begin{equation}
    \psi_{-\uparrow j}=
    \begin{pmatrix}
        i \sqrt{1-\frac{E}{\mu_0}}~K_{j-\frac{1}{2}}(\Bar{r})e^{-i\frac{\pi}{4}+i(j+\frac{n-1}{2})\theta}\\
        0\\
        -i \sqrt{1+\frac{E}{\mu_0}}~K_{j+\frac{1}{2}}(\Bar{r})e^{i\frac{\pi}{4}+i(j-\frac{n-1}{2})\theta}\\
        0
    \end{pmatrix},
\end{equation}  
and the spin-down vortex bound states are expressed as
\begin{equation}
    \psi_{-\downarrow j}=
    \begin{pmatrix}
        0\\
        -i \sqrt{1-\frac{E}{\mu_0}}~K_{j+\frac{1}{2}}(\Bar{r})e^{i\frac{\pi}{4}+i(j-\frac{n-1}{2})\theta}\\
        0\\
        i \sqrt{1+\frac{E}{\mu_0}}~K_{j-\frac{1}{2}}(\Bar{r})e^{-i\frac{\pi}{4}+i(j+\frac{n-1}{2})\theta}
    \end{pmatrix}.
\end{equation} 

Here, $K_\nu(x)~(=K_{-\nu}(x))$ is the modified Bessel function of the second kind.
These functions of neighboring orders are connected by the raising and lowering operators, as given in
\begin{align}
    K_{\nu-1}(x)&=-\qty(\pdv{x}+\frac{\nu}{x})K_\nu(x),\\
    K_{\nu+1}(x)&=-\qty(\pdv{x}-\frac{\nu}{x})K_\nu(x).
\end{align}
The asymptotic behaviors are given by
\begin{equation}
    K_\nu(x)\approx\frac{e^{-x}}{\sqrt{x}}~(x\gg1).
\end{equation}

As required by the single-valuedness of the wave function, $j$ takes integers or half-integers depending on whether the winding number $n$ is odd or even.
Hence, the $j=0$ zero-energy state appears only for odd $n$.
The subscript ''-" in the wave function $\psi_{-\alpha j}~(\alpha=\uparrow\downarrow)$ originates from the fact that the $j=0$ zero-energy state has chirality -1, satisfying $\Gamma\psi_{-\alpha 0}=-\psi_{-\alpha 0}$.

We focus on the low-energy vortex bound states satisfying $E\ll\mu_0$.
Thus, we employ a low-energy effective Hamiltonian for vorex bound states, obtained by setting $r=r_0$ in the angular part [Eq.~\eqref{eq:Ht}] of the original Hamiltonian.
Here $r_0$ represents the characteristic length scale of the vortex, which is on the same order as the coherence length of the superconductor.
In this limit, the expressions for the energy spectrum $E_{-\alpha j}~(\alpha=\uparrow\downarrow)$
are approximated to be
\begin{equation}
    E_{-\uparrow j}\approx-\frac{\Delta_0}{r_0}j,\quad 
    E_{-\downarrow j}\approx\frac{\Delta_0}{r_0}j.\label{eq:energy spectrum of vortex bound state}
\end{equation}
The schematic pictures of the energy spectrum are illustrated in Fig.~\ref{figure:spectrum}.
The spin-up vortex bound states are given by
\begin{equation}
    \psi_{-\uparrow j}\approx g(r)
    \begin{pmatrix}
        ie^{-i\frac{\pi}{4}+i(j+\frac{n-1}{2})\theta}\\
        0\\
        -ie^{i\frac{\pi}{4}+i(j-\frac{n-1}{2})\theta}\\
        0
    \end{pmatrix},
\end{equation}  
and the spin-down vortex bound states are expressed as
\begin{equation}
    \psi_{-\downarrow j}\approx g(r)
    \begin{pmatrix}
        0\\
        -ie^{i\frac{\pi}{4}+i(j-\frac{n-1}{2})\theta}\\
        0\\
        ie^{-i\frac{\pi}{4}+i(j+\frac{n-1}{2})\theta}
    \end{pmatrix},
\end{equation}  
where
\begin{equation}
    g(r)=\frac{1}{\sqrt{r}}\exp\qty[-\frac{\mu_0}{\Delta_0}r]
\end{equation}
In this approximation, all of these states have a definite chirality $\Gamma\psi_{-\alpha j}=-\psi_{-\alpha j}$, not only for the $j=0$ state but also for all $j$.

The above vortex states are not smooth at $r=0$ and it is not clear why they have chirality $\Gamma=-1$, which is opposite to the surface edge states.
In \cite{Aoki:2022cwg, Aoki:2023lqp} 
a similar problem in a systems with a standard magnetic vortex and a monopole was explained by regularizing the short-distance behavior on a lattice.
It was both analytically and numerically shown that the strong curvature at the defects, makes an additive renormalization of the mass term and locally changes the topological phase near the defects.
In this work, we assume that the same mechanism works at a very small but finite radius $r_0$ inside of which $\mu(r)$ goes negative, and $g(r)$ smoothly converges to zero at the origin $r=0$. Then we can identify the above eigenstates as the edge-localized modes of the small domain-wall at $r=r_0$ having the chirality $\Gamma=-1$  and finally neglecting $r_0\to 0$.

The energy spectrum of the vortex bound states ($\Gamma=-1$) is quantized by the unit $\Delta_0/r_0$, 
while the energy spectrum of the edge states ($\Gamma=+1$) is quantized by the unit $\Delta_0/R$.
Therefore, in the limit $r_0 \rightarrow0$, only zero energy state is allowed as a vortex bound state, since the other states are absorbed into the bulk modes.

\section{Zero mode mixing}\label{subsec:zero-mode mixing}
In general, Majorana fermions appear in pairs in finite systems, forming a single complex fermion.
Here, we explicitly demonstrate it in the finite disc geometry, where the hybridization occurs between helical Majorana zero modes at the vortex core and at the edge.
(See also Ref.~\cite{Zhao:2012ue,Aoki:2023lqp}.) 
This hybridization remains robust whatever large separation $R$ between them.


We assume that  the following four functions are good approximations of the original zero modes,
\begin{align}
    \theta(R-r)\psi_{-\uparrow 0},\;\;\; \theta(R-r)\psi_{-\downarrow 0}, \nonumber\\
    \theta(r-r_0)\psi_{+\uparrow 0},\;\;\;  \theta(r-r_0)\psi_{+\downarrow 0}
\end{align}
with the step function $\theta(x)$. We will take $r_0\to 0$ at the end of the computation.
These functions satisfy the appropriate boundary conditions at the edge $r=R$ and at the vortex $r=r_0\to 0$ when $-\mu(r)$ is sufficiently large for $r>R$, and $r<r_0$.

The above approximated zero modes are no more eigenstates of the original Hamiltonian $H_\mathrm{BdG}=H_r+H_\theta$.
But we can assume that the true eigenmodes are well approximated by 
linear combinations of them. In order to solve this problem, 
let us compute the matrix elements of $H_\mathrm{BdG}$ among these states.
Noting that the $J$  operation is trivially zero,  we have
\begin{align}
    H_\mathrm{BdG} \theta(R-r)\psi_{-\alpha 0} &= \Delta_0(\sigma_3\otimes 1)\delta(R-r) \psi_{-\alpha 0} ,\nonumber\\
    H_\mathrm{BdG} \theta(r-r_0)\psi_{+\alpha 0}  &= \Delta_0(\sigma_3\otimes 1)\delta(r-r_0)\psi_{+\alpha 0},
\end{align}
for each $\alpha=\uparrow\downarrow$ and the matrix elements in the $r_0\to 0$ limit are
\begin{align}
  &\int_0^\infty drr \int_0^{2\pi}d\theta [\theta(r-r_0)\psi_{+\alpha 0}]^\dagger H \theta(R-r)\psi_{-\beta 0}\nonumber\\
  &= -\int_0^\infty drr \int_0^{2\pi}d\theta [\theta(R-r)\psi_{-\alpha 0}]^\dagger H \theta(r-r_0)\psi_{+\beta 0}
  \nonumber\\
  &\to_{r_0\to 0}i \delta_{\alpha\beta}\epsilon,
\end{align}
where we have defined $\delta_{\uparrow\uparrow}=\delta_{\downarrow\downarrow}=1,\delta_{\uparrow\downarrow}=\delta_{\downarrow\uparrow}=0 $ and 
\begin{align}
        \epsilon &= 4\pi \Delta_0Rf(R)g(R) = 4\pi \Delta_0\lim_{r_0\to 0}r_0f(r_0)g(r_0)\nonumber\\
        &=4\pi \Delta_0\exp\left[-\frac{1}{\Delta_0}\int_0^R dr' \mu(r')\right].
\end{align}
The other matrix elements are all zero.

Although $\epsilon$ is exponentially small, the Hamiltonian
has an off-diagonal substructure for each $\alpha=\uparrow\downarrow$ 
\begin{equation}
    H=\left(\begin{array}{cc}
    0 & i\epsilon\\
    -i\epsilon & 0
    \end{array}\right)
\end{equation}
and the true eigenfunctions are maximally mixed:
\begin{equation}
\label{eq:zero-mix}
\psi_\alpha^{\pm \epsilon}=\frac{1}{\sqrt{2}}\lim_{r_0\to 0}\left[
    \theta(R-r)\psi_{-\alpha 0}
   \mp i \theta(r-r_0)\psi_{+\alpha 0}\right],
\end{equation}
with the eigenvalues $\pm \epsilon$.
Note that the mixing between the edge mode and the vortex mode 
persists with whatever large value of $R$.
Besides, $\psi_\alpha^{\pm \epsilon}$ are not eigenstates of
$C$ but interchange as
\begin{equation}
    C\psi_\alpha^{\pm \epsilon}=\psi_\alpha^{\mp \epsilon}.
\end{equation}
Therefore, the hybridization between the two distinct Majorana zero modes remains robust irrespective of the separation between the zero-energy bound states, no matter how far apart they are, even in the limit $R\rightarrow\infty$.

The above analysis is completely parallel to the case of the Majorana zero modes that appear in $(p_x\pm ip_y)$-chiral superconductors.
Thus, the non-local complex fermion given in Eq.~(\ref{eq:zero-mix}), that is made up with localized Majorana fermions,
can be exploited, for example, to realize non-Abelian braiding statistics~\cite{Ivanov2001,Sato2003}, thereby enabling potential applications in topological quantum computation.

The discussion of zero mode mixing is also valid in a three-dimensional system [Sec.~\ref{subsec:extension to 3D}], where the time-reversal invariant vortex becomes one dimensional defect.
In a cylindrical geometry with open boundaries, when a straight vortex line is positioned along the cylinder’s axis, the helical Majorana zero modes localized at the vortex core and on the outer surface hybridize to form a non-local complex fermion.
On the other hand, in a closed system where a pair of ring-shaped vortices are linked, as depicted in the Fig. \ref{figure:vortex_in_3D}(b), the helical Majorana zero modes localized on each vortex ring combine to form a non-local complex fermion.




\bibliography{reference}

\end{document}